\documentclass[twocolumn,showkeys,showpacs,preprintnumbers,prd,superscriptaddress,nofootinbib]{revtex4-2}

\usepackage{graphicx,amsmath}
\usepackage{dcolumn}
\usepackage{bm}
\usepackage[dvipsnames]{xcolor}
\usepackage{multirow}
\usepackage{journals}
\usepackage{verbatim}
\usepackage[T1]{fontenc}
\usepackage[utf8]{inputenc}
\usepackage{booktabs}
\usepackage[table]{xcolor}
\usepackage{color}
\usepackage{transparent}
\usepackage{pifont}
\usepackage{enumitem,float}

\usepackage{hyperref}
\hypersetup{
    colorlinks=true, 
    linkcolor=blue, 
    citecolor=magenta}

\begin{document}


\title{$\mathbf{\Omega_1\Omega_2}$--$\mathbf{\Lambda}$CDM: A promising phenomenological extension of the standard model of cosmology}%

\author{Suresh Kumar}
\email{suresh.kumar@plaksha.edu.in}
\affiliation{Plaksha University, Mohali, Punjab-140306, India}


\begin{abstract}
We investigate a phenomenological extension of the standard $\Lambda$CDM framework, the $\Omega_1\Omega_2$-$\Lambda$CDM model, in which the total energy density of the universe is expanded in powers of $1+z$. This parameterization recovers the standard $\Lambda$CDM scenario and  contributes two additional terms, $\Omega_1(1+z)$ and $\Omega_2(1+z)^2$, in the dark energy sector, alongside the cosmological constant, leading to a physically interpretable and observationally testable dynamics of effective dark energy. Using Planck cosmic microwave background (CMB) data, we find that the model allows increased freedom in the late-time expansion history while preserving the standard early-universe physics. When DESI baryon acoustic oscillation (BAO) data are included, the inferred Hubble constant is $H_0 = 69.74 \pm 0.77~\mathrm{km\,s^{-1}\,Mpc^{-1}}$ (68\% CL), which is consistent with recent Tip of the Red Giant Branch (TRGB) measurements from the Carnegie--Chicago Hubble Program (CCHP). This agreement indicates internal consistency between Planck CMB, DESI BAO, and TRGB-based measurements across a wide redshift range. The reconstructed dark energy equation of state exhibits a smooth transition across the phantom divide followed by asymptotic de Sitter behavior, leading to modified late-time dynamics of universe while maintaining standard early-time cosmology. Overall, our results demonstrate that controlled late-time deviations from $\Lambda$CDM can improve cosmological concordance.
\end{abstract}
\maketitle

\section{Introduction}
\label{sec:introduction}

The standard cosmological model, $\Lambda$CDM, provides a remarkably successful framework for understanding the evolution of our universe from its earliest moments to the present day \cite{Weinberg:2008zzc}. It is well-supported by an extensive array of observational evidence from the precise measurements of the cosmic microwave background (CMB) anisotropies by WMAP \cite{WMAP:2003elm, WMAP:2007kom} and Planck \cite{Planck:2018vyg} to the large-scale distribution of galaxies and thereby considered as the cornerstone of modern cosmology. The model carries the assumption of a cosmological constant $\Lambda$, representing a constant energy density that drives the accelerated expansion of the universe at late times  and cold dark matter (CDM) that governs structure formation \cite{Peebles:1984ge}.

Increasingly precise cosmological observations have revealed persistent discrepancies within the standard $\Lambda$CDM cosmology, most notably in measurements of the present-day Hubble expansion rate, $H_0$. The so-called \textit{Hubble tension} refers to the disagreement between early-universe determinations of $H_0$, inferred primarily from the cosmic microwave background (CMB) observations under the assumption of $\Lambda$CDM, and late-universe measurements based on the local distance ladder
\cite{Perivolaropoulos:2021jda,Abdalla:2022yfr,DiValentino:2021izs,Verde:2019ivm,Schoneberg:2021qvd,Vagnozzi:2023nrq,Perivolaropoulos:2024yxv,Verde:2023lmm,CosmoVerseNetwork:2025alb}.
Early-universe constraints from the Planck satellite CMB measurements, supported by complementary ground-based CMB experiments like the Atacama Cosmology Telescope (ACT) and the South Pole Telescope – 3rd Generation Camera (SPT-3G), consistently favor a lower value of $H_0$ near $67\ \mathrm{km\,s^{-1}\,Mpc^{-1}}$ when interpreted within $\Lambda$CDM \cite{Planck:2018vyg,SPT-3G:2025bzu}. In contrast, the late-universe measurements from the SH0ES (Supernovae and $H_0$ for the Equation of State) collaboration, based on a Cepheid-calibrated Type Ia supernova distance ladder, yield a significantly higher value of $H_0$ around $73\ \mathrm{km\,s^{-1}\,Mpc^{-1}}$ \cite{Riess:2021jrx,Breuval:2024lsv}. Recently, a similar high value of $H_0$ is reported by Local Distance Network \cite{H0DN:2025lyy}. The statistical discrepancy between these early- and late-time determinations exceeds $5-7\sigma$ level, making it one of the most severe internal tensions in modern cosmology. 

An important development in this debate is the emergence of independent local measurements of $H_0$ based on the Tip of the Red Giant Branch (TRGB) approach \cite{Freedman:2024eph}. This technique relies on Population II red giant stars in the helium-burning phase, predominantly observed in the halos of nearby galaxies where stellar crowding and dust extinction are minimal. The characteristic luminosity increase at the helium flash, the TRGB, is well understood from stellar evolution theory and exhibits only a weak, calibratable dependence on metallicity, making it a robust standard candle \cite{madore2023population}. Recent TRGB-based calibrations, anchored to geometric distance indicators (e.g., detached eclipsing binaries, Milky Way parallaxes, or masers) and applied to Type Ia supernovae, consistently find values of $H_0$ around $69$–$70\ \mathrm{km\,s^{-1}\,Mpc^{-1}}$ \cite{Freedman:2019jwv,Freedman:2024eph}. This intermediate positioning of TRGB-based $H_0$ measurements, sitting between the Planck-inferred value and the SH0ES measurement, has important implications for interpreting the Hubble tension.
Furthermore, the TRGB collaboration under the Carnegie-Chicago Hubble Program (CCHP) \cite{Freedman:2024eph} argues that the statistical significance of the Hubble tension may be overstated. They note that successive SH0ES analyses have shifted by ~1.6\% due to newly identified systematics, comparable to the total quoted uncertainty, and that these correlated updates violate the assumption of statistical independence used in tension estimates. This suggests that the true uncertainty in local $H_0$ measurements may be significantly different from the currently reported ones, and that claims of a $5\sigma$ discrepancy with $\Lambda$CDM may be premature until representative systematic uncertainties are fully accounted for \cite{Freedman:2024eph}.

While the SH0ES result alone strongly favors new physics beyond $\Lambda$CDM and TRGB result prefers a middle ground, the potential for residual, correlated systematics remains a topic of active debate \cite{Efstathiou:2021ocp,Hogas:2024qlt}. The current observational landscape therefore does not yet provide a definitive prediction of the Hubble tension, but it does motivate the exploration of cosmological models that allow controlled departures from the standard late-time expansion history while preserving early-universe physics. Phenomenological extensions of $\Lambda$CDM offer a flexible and data-driven approach to testing whether modest late-time dynamics of dark energy can reconcile the diverse set of cosmological and local measurements without invoking drastic modifications to the early-universe physics \cite{CosmoVerseNetwork:2025alb}. 

In this work, we adopt a \textit{reconstruction-first} approach leading to a phenomenological extension of $\Lambda$CDM. Rather than postulating a specific microphysical model for dark energy, we parameterize the total energy density of the universe as a phenomenological but mathematically and physically motivated expansion in powers of $(1+z)$, building upon phenomenological approaches explored in \cite{Sahni:2002fz,Alam:2004jy, Linder:2006xb} and reconstruction methods in \cite{Holsclaw2010, Shafieloo2012}. This leads to the $\Omega_1\Omega_2$--$\Lambda$CDM model, where deviations from a pure cosmological constant are captured by two additional parameters: $\Omega_1$ and $\Omega_2$, which correspond to terms scaling as $(1+z)$ and $(1+z)^2$, respectively. This framework is minimal and flexible, allowing for a wide range of late-time expansion histories. The cosmic evolution in this model at early times remain largely unaffected, preserving the successful predictions of early universe physics, while enabling substantial freedom at low redshifts where the Hubble tension manifests.

We identify transient phantom behavior of the $\Omega_1\Omega_2$--$\Lambda$CDM model, and finally  proceed for the observational analysis of the model in two stages. First, we constrain the $\Omega_1\Omega_2$--$\Lambda$CDM model using Planck CMB data alone, and demonstrate that the geometric degeneracy inherent in CMB observations \cite{Efstathiou:1998xx} allows for significantly higher values of $H_0$ when late-time restrictions are relaxed. Second, we use the latest baryonic acoustic oscillation (BAO) measurements from the Dark Energy Spectroscopic Instrument (DESI) \cite{DESI:2025zgx}, which provide crucial anchoring of the expansion history at intermediate redshifts ($0.3 \lesssim z \lesssim 2.3$). Their combination with Planck CMB data breaks degeneracies and yields tight constraints on the model parameters. We show that this approach alleviates the Hubble tension predicted by TRGB measurements. The model predicts a smooth, non-monotonic evolution of the dark energy, featuring a transient phantom phase that enhances late-time acceleration. Remarkably, this is achieved without altering the early-universe parameters, highlighting the power of reconstruction-first methods for late-time modifications.

This paper is structured as follows. In Section~\ref{sec:model}, we provide the theoretical framework of the $\Omega_1\Omega_2$--$\Lambda$CDM model and derive its key background equations.   The observational datasets, evolution of linear perturbations and statistical methodology employed in our analysis are described in Section~\ref{sec:data_methods}.  Main results and discussion of the observational constraints on model parameters are reported in Section~\ref{sec:results}. A comparison of the model and results with some existing dynamical dark energy models is presented in Section \ref{sec:comparison}.  Finally, section~\ref{sec:conclusions} summarizes main findings of this study with some future directions.

\section{The $\Omega_1\Omega_2$--$\Lambda$CDM model} \label{sec:model}

The dynamics of a spatially flat, homogeneous, and isotropic universe is governed by the Friedmann equation:
\begin{equation}\label{frd}
H^2(z) = \frac{8\pi G}{3} \rho_{\rm tot}(z),
\end{equation}
where $H(z)$ is the Hubble parameter, $\rho_{\rm tot}(z)$ is the total energy density and $G$ is Newton's gravitational constant. Observational data (e.g., from BAO, supernovae and CMB) constrain $H(z)$ directly. So, a parameterization of $\rho_{\rm tot}(z)$ provides the most straightforward link between theory and observations. It is therefore worth investigating a well-motivated parametrization of $\rho_{\rm tot}(z)$.\\

In this work, we consider a truncated Taylor expansion of the total energy density of universe in powers of $1+z$ \cite{Sahni:2002fz,Alam:2004jy}:
\begin{equation}
\label{eq:rho_series}
\rho_{\rm tot}(z) = c_0 + c_1 (1+z) + c_2 (1+z)^2 + c_3 (1+z)^3 + c_4 (1+z)^4.
\end{equation}
This is not a generic Taylor expansion\footnote{Although this form resembles a Taylor expansion, it may instead be interpreted as a global polynomial approximation of the total energy density over a redshift interval probed by the observations. By the classical Weierstrass approximation theorem, any continuous function on a closed interval can be uniformly approximated arbitrarily well by polynomials, regardless of the convergence properties of its Taylor series \cite{Rudin1976,Cheney1982}.} to provide a genuine mathematical approximation of the true energy density function of the universe but a deliberate choice with direct physical significance. For, each term $\rho_n(z) \propto (1+z)^n$ corresponds exactly to the evolution of a perfect fluid with a constant equation of state parameter:
$w_n = \frac{n}{3} - 1$. The first term $c_0$, which is independent of the redshift, corresponds to a cosmological constant or vacuum energy with $w=-1$. The linear scaling term $c_1 (1+z)$ is associated with an exotic component with $w=-2/3$, often linked to domain walls, topological defects, or an effective dark energy contribution beyond $\Lambda$. The quadratic term $c_2 (1+z)^2$ behaves as a fluid with $w=-1/3$ and is commonly interpreted as an effective curvature-like contribution. The cubic term $c_3 (1+z)^3$ reproduces the standard scaling of pressureless and non-relativistic matter with $w=0$, while the quartic term $c_4 (1+z)^4$ with $w=1/3$ corresponds to radiation and other relativistic species. This decomposition of $\rho_{\rm tot}(z)$ provides an elegant framework for investigating departure from $\Lambda$CDM onto physically-interpretable components.

Next, retaining terms up to $n=4$ ensures the correct description of background evolution of the universe from the radiation-dominated era ($n=4$) to the present. The lower-order terms ($n=0,1,2$) considered together constitute an effective dark energy sector, which can drive late-time acceleration of the universe without committing to a specific microphysical origin. Further, the choice of $(1+z) = a^{-1}$, where $a$ is scale factor, as the expansion variable has a compelling mathematical motivation: it represents an expansion around the infinite future ($a \to \infty$, $z \to -1$). In dynamical systems language, the asymptotic future is an infrared fixed point of the cosmic expansion \cite{Wald1983,Copeland:1997et}. An expansion in $(1+z)$ is thus analogous to a Taylor series expansion around this attractor, a common technique in effective field theory \cite{Cheung:2007st,Gubitosi:2012hu}. As $z \to -1$, all terms with $n \ge 1$ vanish, leaving $\rho_{\rm tot} \to c_0$. Thus, the universe asymptotically approaches a de Sitter state in this framework. Also, it guarantees a future evolution free of singularities, and consistent with the cosmic ``no-hair'' conjectures \cite{Wald1983}. 

At the background level, different physical components may share the same redshift scaling. For instance, the $c_3 (1+z)^3$ term combines both baryons and cold dark matter. The $c_2 (1+z)^2$ term can represent either the spatial curvature ($k \neq 0$) in a non-flat universe or an exotic dark energy fluid with $w=-1/3$, or a combination thereof. The parametrization considered here does not artificially break this degeneracy, and therefore the terms represent the total contribution scaling as $(1+z)^{n}$. So it is an overall measure of the expansion history $H(z)$. In order to distinguish between microphysical origins, it requires additional non-geometrical data (e.g., growth of structure, CMB anisotropies).

After identifying the energy components of the universe in Eq.~\eqref{eq:rho_series}, the Friedmann equation~\eqref{frd} may now be expressed in terms of the standard dimensionless density parameters as
\begin{equation}
\label{eq:H2_final}
\frac{H^2(z)}{H_0^2}
= \Omega_{\mathrm{r}} (1+z)^4
+ \Omega_{\mathrm{m}} (1+z)^3
+ \Omega_{\Lambda}
+ \Omega_1 (1+z)
+ \Omega_2 (1+z)^2 ,
\end{equation}
which explicitly shows the redshift dependence of each energy component contributing to the expansion of universe. Assuming spatial flatness, these density parameters satisfy the normalization condition: $\Omega_{\mathrm{r}} + \Omega_{\mathrm{m}} + \Omega_{\Lambda} + \Omega_1 + \Omega_2 = 1$. All deviations from standard $\Lambda$CDM (where $\Omega_1=\Omega_2=0$) are absorbed into or captured by the dynamical terms, $\Omega_1(1+z)$ and $\Omega_2(1+z)^2$. It may be noted that the baryonic and cold dark matter co-exist in the pressureless matter sector with the same scaling $(1+z)^3$, and are measured separately by the observations. Likewise, the dark energy contribution $\Omega_2(1+z)^2$ may be distinguished from the curvature term $\Omega_k(1+z)^2$ by the observations, and the same applies to the other dark energy term $\Omega_1(1+z)$. Here, we attribute the terms $\Omega_1(1+z)$ and $\Omega_2(1+z)^2$ along with $\Omega_{\Lambda}$ to the effective dark energy sector .  Overall, the parametrization adopted here is both minimal and complete (encompassing all constant-$w$ fluids up to radiation) for describing cosmic evolution from recombination to the de Sitter epoch. For convenience, we refer this framework to as the $\Omega_1\Omega_2$--$\Lambda$CDM model.

The effective dark energy density is the sum of the lower-order terms \cite{Alam:2004jy}, given by
\begin{equation}
\rho_{\rm de}(z) =\rho_{\rm crit, 0} [\Omega_{\Lambda} + \Omega_1 (1+z) + \Omega_2 (1+z)^2],
\end{equation}
where $\rho_{\rm crit, 0}=\frac{3H_0^2}{8\pi G}$ is the present-day critical density. From the continuity equation, $\dot{\rho}_{\rm de} + 3H(1+w_{\rm de})\rho_{\rm de}=0$ (overdot indicating the cosmic time derivative), we derive the effective equation of state of dark energy:
\begin{equation}
\label{eq:weff}
w_{\rm de}(z) = -1 + \frac{1}{3} \frac{\Omega_1 (1+z) + 2\Omega_2 (1+z)^2}{\Omega_{\Lambda} + \Omega_1 (1+z) + \Omega_2 (1+z)^2}.
\end{equation}
This $w_{\rm de}(z)$ is a rational function, allowing for rich dynamics of dark energy including transitions across the phantom divide ($w = -1$). We find $w_{\rm de}(z \to -1) = -1 \quad \text{(de Sitter attractor)}$ and $w_{\rm de}(z \to \infty) = -\frac{2}{3} \quad \text{if } \Omega_2 = 0, \text{ or } -\frac{1}{3} \quad \text{if } \Omega_2 \neq 0$. This limiting behavior shows that the model naturally avoids pathological future singularities. Also, the conditions $\Omega_\Lambda>0$, $\Omega_2>0$ and $\Omega_1^2<4\Omega_2\Omega_\Lambda$ are sufficient to ensure non-singular behavior of $w_{\rm de}$ for $z\geq -1$. Thus, $\Omega_1<0$ is permissible in this model which may lead to $w_{\rm de}<-1$. Such a behavior of $\Omega_1\Omega_2$--$\Lambda$CDM model may be realized in a Quintom model consisting of two scalar fields \cite{Vikman:2004dc,Copeland:2006wr,Feng:2004ad,Guo:2004fq,Cai:2009zp}. 

In the following section, we test the $\Omega_1\Omega_2$--$\Lambda$CDM model with the observational data and see whether Quintom behavior is preferred by the data within this framework. 

\section{Observational Data Analysis}\label{sec:data_methods}
\subsection{Datasets}

For observational data analysis of the $\Omega_1\Omega_2$--$\Lambda$CDM model, first we employ CMB data from the Planck 2018 legacy release in order to constrain early-universe physics and the angular diameter distance to recombination in the model. Specifically, the temperature anisotropy and polarization power spectra measurements, their cross-spectra, and the lensing power spectrum~\cite{Planck:2019nip,Planck:2018lbu} are used which include the high-$\ell$ \texttt{Plik} likelihood for TT ($30 \leq \ell \leq 2508$) along with TE and EE ($30 \leq \ell \leq 1996$), the low-$\ell$ TT-only likelihood ($2 \leq \ell \leq 29$) based on the \texttt{Commander} component-separation algorithm, the low-$\ell$ EE-only likelihood ($2 \leq \ell \leq 29$) using the \texttt{SimAll} method, and measurements of CMB lensing. We refer this combined CMB dataset to as Planck.

It deserves mention that the CMB tightly constrains $\Lambda$CDM model but not the extended dark-energy models. For, the CMB anisotropies are produced primarily at the epoch of recombination,
$z_\ast \simeq 1100$, when the universe was mainly radiation and matter dominated. In the standard
$\Lambda$CDM scenario, the dark energy density is negligible at this epoch, $\Omega_\Lambda(z_\ast) \ll 1$, thereby the expansion history and perturbation dynamics at recombination are dictated almost entirely by the
physical matter and baryon densities, $\omega_{\mathrm{m}} \equiv \Omega_{\mathrm{m}} h^2$ and $\omega_b \equiv \Omega_b h^2$,
along with the radiation content. As a consequence, the CMB alone provides high-precision measurements of the physical matter and baryon densities
($\omega_{\mathrm{m}}$, $\omega_b$), the primordial power-spectrum parameters $(A_s,n_s)$, and the sound horizon at
recombination, $r_s(z_\ast)$, via the detailed structure of the acoustic peaks and the damping tail.

The most precisely measured geometric quantity by the CMB is the angular acoustic scale \cite{Planck:2018vyg},
$\theta_\ast = \frac{r_s(z_\ast)}{D_A(z_\ast)}$, where $D_A(z_\ast)$ is the comoving angular diameter distance to the last scattering,
$D_A(z_\ast) = \int_0^{z_\ast} \frac{dz}{H(z)}$.
The sound horizon $r_s(z_\ast)$ depends almost exclusively on pre-recombination physics and is therefore
fixed by the physical densities $\omega_{\mathrm{m}}$ and $\omega_b$, whereas $D_A(z_\ast)$ is sensitive to the integrated expansion history of the universe
from $z=0$ to $z_\ast$.  This limited freedom in $H(z)$ within the $\Lambda$CDM framework implies that $\theta_\ast$ strongly
constrains the model parameters, rendering the $\Lambda$CDM scenario effectively over-constrained by the CMB data alone.

On the other hand, extensions of $\Lambda$CDM that introduce additional dark energy degrees of freedom such as
a redshift-dependent equation of state $w(z)$ or extra density parameters $(\Omega_1,\Omega_2)$ in $\Omega_1\Omega_2$--$\Lambda$CDM model, are
typically expected to satisfy $\rho_{\rm de}(z \gtrsim 10^3) \ll \rho_{\rm m}, \rho_{\rm r}$, ensuring consistency with early-universe physics. Consequently, these additional parameters of background evolution exhibit a
negligible impact on the primary CMB anisotropies and enter the CMB observables mainly through the distance
integral, $D_A(z_\ast)$. Notably, many different late-time expansion histories can yield the same
$D_A(z_\ast)$, and therefore the CMB alone exhibits a pronounced geometric degeneracy in the presence of extended dark-energy sectors.

This degeneracy manifests most clearly in correlations with the matter density and Hubble constant.
It deserves mention that the CMB data tightly constrain the physical matter density $\omega_{\mathrm{m}} = \Omega_{\mathrm{m}} h^2$, but not $\Omega_{\mathrm{m}}$ and
$H_0$ separately. Therefore, modifications to the late-time dark energy sector can alter $H_0$ while
compensating changes in $\Omega_{\mathrm{m}}$ in order to preserve $\theta_\ast$, leading to strong degeneracies in the
$(\Omega_{\mathrm{m}},H_0)$ plane. Similarly, CMB accurately measures the primordial amplitude
$A_s e^{-2\tau}$ and the growth of perturbations up to recombination but the late-time growth history depends
on the dark energy evolution. As a result, parameters such as $S_8$ also become weakly constrained and highly
model-dependent when additional dark energy freedom is introduced in the expansion history.

Further, the late-time CMB effects, such as the integrated Sachs--Wolfe (ISW) effect and CMB lensing, offer only
limited additional constraining power because the ISW effect is confined to large angular scales and is strongly
cosmic-variance limited, while CMB lensing probes an integrated combination of growth and geometry and thereby
remains degenerate with other parameters, including neutrino masses and $A_s$. So, these CMB probes
are insufficient to fully break the degeneracies associated with extended dark energy model parameters.

Thus, the apparent preference of the CMB data for the $\Lambda$CDM scenario reflects the fact that the CMB is primarily
an early-universe experiment. Any dark energy model that closely mimics $\Lambda$CDM model at the high redshifts but
deviates only at late times can fit the CMB data comparably well, at the cost of introducing strong
degeneracies with $H_0$, $\Omega_{\mathrm{m}}$, and growth-related parameters. Robust constraints on additional
dark energy parameters by breaking the CMB-induced degeneracies therefore require the inclusion of complementary low-redshift probes, such as
BAO, Type~Ia supernovae, redshift-space distortions, weak lensing and direct
measurements of $H(z)$.

In this work, we incorporate BAO distance measurements from the Dark Energy Spectroscopic Instrument (DESI). These measurements constrain the transverse and radial distance scales, $D_M(z)/r_{\rm d}$ and $D_H(z)/r_{\rm d}$, thereby probing the redshift evolution of the expansion history. The BAO measurements from the DESI BAO DR2 release~\cite{DESI:2025zgx} include galaxy, quasar, and Lyman-$\alpha$ forest tracers over the redshift range $0.1< z<4.6$, divided into seven redshift bins, with the effective redshift range $0.295\leq z\leq2.330$. The constraints are expressed in terms of $D_{\mathrm{M}}/r_{\mathrm{d}}$, $D_{\mathrm{H}}/r_{\mathrm{d}}$, and $D_{\mathrm{V}}/r_{\mathrm{d}}$, normalized by $r_{\mathrm{d}}$, the sound horizon at the drag epoch. We refer this dataset to as DESI. 

\subsection{Linear Perturbations}
\label{sec:linear_perturbations}

In the synchronous gauge comoving with CDM \cite{Ma:1995ey},
the line element for scalar perturbations reads
\begin{equation}
ds^2 = a^2(\tau)\left[-d\tau^2 + (\delta_{ij}+h_{ij})dx^i dx^j \right],
\end{equation}
where $\tau$ denotes conformal time and $h_{ij}$ represents the metric
perturbations. The residual gauge freedom of the synchronous gauge is fixed
by choosing the rest frame of CDM, $\theta_c=0$. The evolution of perturbations
is governed by the coupled Einstein--Boltzmann equations, which we solve
numerically using the \texttt{CLASS} code \cite{Blas:2011rf}.

For a general fluid component $i$ with equation-of-state parameter $w_i$
and rest-frame sound speed $c_{s,i}^2$, the density contrast
$\delta_i\equiv\delta\rho_i/\bar{\rho}_i$ and the velocity divergence
$\theta_i$ obey the linearized continuity and Euler equations in Fourier space, given by
\begin{align}
\delta_i' &=
-(1+w_i)\left(\theta_i + \frac{h'}{2}\right)
- 3\mathcal{H}\left(c_{s,i}^2-w_i\right)\delta_i, \\
\theta_i' &=
-\mathcal{H}\left(1-3w_i\right)\theta_i
- \frac{w_i'}{1+w_i}\theta_i
+ \frac{c_{s,i}^2}{1+w_i}k^2\delta_i
- k^2\sigma_i ,
\label{eq:pert_eqs}
\end{align}
where primes
denote derivatives with respect to conformal time, $\mathcal{H}\equiv a'/a$ is the conformal Hubble parameter, $k$ is the
comoving wavenumber, and $\sigma_i$ denotes the anisotropic stress. For perfect fluids,
$\sigma_i=0$. For CDM ($w_c=0$, $c_{s,c}^2=0$) and baryons
($w_b=0$, $c_{s,b}^2\neq0$), the above perturbation equations reduce to the standard
fluid equations implemented in \texttt{CLASS}. Radiation perturbations
($w_r=1/3$) are treated using the full Boltzmann hierarchy for photons
and neutrinos.

For adiabatic dark energy perturbations, the pressure perturbation is
determined by the adiabatic sound speed
\begin{equation}
c_{a,\rm{de}}^2 \equiv \frac{p_{\rm de}'}{\rho_{\rm de}'}
= w_{\rm de}
- \frac{1+z}{3\left(1+w_{\rm de}\right)}\frac{dw_{\rm de}}{dz},
\label{eq:adiabatic_sound}
\end{equation}
which follows from energy--momentum conservation. In this work, we neglect
non-adiabatic (entropy) perturbations.

For the $\Omega_1\Omega_2$-$\Lambda$CDM model, $c_{a,\rm{de}}^2$ becomes negative over a wide redshift range and diverges at the phantom-divide crossing where $1+w_{\rm de} \to 0$. This divergence renders the standard fluid equations \eqref{eq:pert_eqs} ill-defined, as both terms $\frac{w_{\rm de}'}{1+w_{\rm de}}\theta_{\rm de}$ and $\frac{c_{s,\rm de}^2}{1+w_{\rm de}}k^2\delta_{\rm de}$ become singular. This reflects the breakdown of a purely adiabatic fluid description rather than a physical instability.


To overcome this limitation, we implement dark-energy perturbations using the parametrized post-Friedmann (PPF) formalism \cite{Fang:2008sn,Hu:2008zd} as available in \texttt{CLASS}. The PPF framework provides a stable and gauge-invariant description of dark-energy perturbations across the phantom divide by replacing the ill-defined fluid variables with a single auxiliary degree of freedom that interpolates consistently between super-horizon and sub-horizon scales. In this framework, we assume a unit rest-frame sound speed, $c_s^2 = 1$, which ensures the absence of gradient instabilities and corresponds to the standard assumption for smooth dark energy. Via the PPF formalism, matter and metric perturbations are evolved self-consistently, with dark energy contributing both through the background expansion rate and through its perturbative effects on large scales.


The impact of dark-energy perturbations is confined to late cosmic times and to large cosmological scales ($k \lesssim 0.01\,h\,\mathrm{Mpc}^{-1}$ or $z \lesssim 2$), where dark energy contributes non-negligibly to the total energy density and influences the evolution of the gravitational potentials. This primarily affects the late-time Integrated Sachs--Wolfe effect and large-scale growth observables such as $f\sigma_8$. On sub-horizon scales, dark-energy perturbations are strongly suppressed due to pressure support, rendering their effect negligible for small-scale clustering observables such as galaxy-galaxy lensing or redshift-space distortions on BAO scales  \cite{Weller:2003hw,Bean:2003fb,Hu:2004yd,Sapone:2009mb}. 

Thus, the PPF framework allows us to consistently and robustly include dark-energy perturbations in a model that exhibits phantom-divide crossings, ensuring theoretical stability and reliable comparison with cosmological observations while preserving the model's predictive power for background observables. A full perturbation analysis and microphysical embedding (e.g., Quintom realizations \cite{Vikman:2004dc,Copeland:2006wr,Feng:2004ad,Guo:2004fq,Cai:2009zp}) are planned for future work.

\subsection{Codes and Numerical Settings}
We implement the theoretical models under consideration in the \texttt{CLASS} Boltzmann solver~\cite{Blas:2011rf} and perform Monte Carlo analyses using the \texttt{MontePython} sampler~\cite{Audren:2012wb, Brinckmann:2018cvx}. Markov chains convergence is ensured by the Gelman-Rubin criterion~\cite{Gelman_1992} with $R-1\leq10^{-2}$, and we process the chains using the \texttt{GetDist} package for posterior extraction\footnote{\url{https://github.com/brinckmann/montepython_public}} and statistical summaries. We assume three degenerate species of massive neutrinos with minimal masses consistent with oscillation experiments with the number of species: $N_{\mathrm{eff}} = 3.044$ (including non-thermal corrections from neutrino decoupling) and mass hierarchy: One massive neutrino with $m_{\nu} = 0.06$~eV and two massless species. The parameter space of the $\Omega_1\Omega_2$--$\Lambda$CDM model includes the  usual six standard parameters of the $\Lambda$CDM model: the present-day physical densities of baryons, $\omega_{\rm b} \equiv \Omega_{\rm b}h^2$, and cold dark matter, $\omega_{\rm cdm} \equiv \Omega_{\rm cdm}h^2$; the angular size of the sound horizon at recombination, $\theta_{\rm s}$; the amplitude of primordial scalar perturbations, $\ln(10^{10}A_{\rm s})$; the scalar spectral index, $n_{\rm s}$; and the optical depth to reionization, $\tau_{\rm reio}$, in addition, the parameters $\Omega_1$ and $\Omega_2$ that characterize the effective dark energy. Flat priors are adopted for all parameters: $\omega_{\rm b} \in [0.018, 0.024]$, $\omega_{\rm cdm} \in [0.10, 0.14]$, $100\theta_{\rm s} \in [1.03, 1.05]$, $\ln(10^{10}A_{\rm s}) \in [3.0, 3.18]$, $n_{\rm s} \in [0.9, 1.1]$, $\tau_{\rm reio} \in [0.04, 0.125]$, $\Omega_1 \in [-1, 0.5]$ and $\Omega_2 \in [0, 0.1]$. Here, the positive prior range of $\Omega_2$ is in line with the earlier observation that $\Omega_2>0$ is essential to avoid non-singular behavior of $w_{\rm de}$. These choices therefore restrict the parameter space to regions where the dark energy equation of state is well behaved and prevent the exploration of unphysical solutions. More wider, unconditioned priors can lead to strong prior-volume effects and parameter sensitivities, particularly in Planck CMB-only analyses. The adopted priors here therefore provide a more physically meaningful and conservative framework for exploring departures from $\Lambda$CDM within the $\Omega_1\Omega_2$--$\Lambda$CDM model.

\begin{table}[H]
\caption{Marginalized constraints (68\% CL) for the baseline and some derived parameters of  $\Omega_1\Omega_2$-$\Lambda$CDM and $\Lambda$CDM models using the Planck and Planck+DESI datasets. The $\Lambda$CDM constraints are shown in blue. The table also includes $\Delta \chi^2_{\text{min}} = \chi^2_{\text{min ($\Omega_1\Omega_2$-$\Lambda$CDM)}} - \chi^2_{\text{min ($\Lambda$CDM)}}$, where negative values indicate a better fit for the $\Omega_1\Omega_2$-$\Lambda$CDM compared to the $\Lambda$CDM model.}
\label{tab:results}
\centering
\begin{tabular}{l l l}
\hline
Data & Planck & Planck+DESI \\
\hline
Model & $\Omega_1\Omega_2$-$\Lambda$CDM & $\Omega_1\Omega_2$-$\Lambda$CDM \\
 & {\color{blue}$\Lambda$CDM} & {\color{blue}$\Lambda$CDM} \\
\hline
$10^{2}\omega{}_{\rm b}$ & $2.242\pm 0.015$ & $2.247\pm 0.013$ \\
 & {\color{blue}$2.240\pm 0.014$} & {\color{blue}$2.255\pm 0.013$} \\[2pt]
$\omega{}_{\rm cdm}$ & $0.1194\pm 0.0013$ & $0.11872\pm 0.00093$ \\
 & {\color{blue}$0.1198\pm 0.0012$} & {\color{blue}$0.11773\pm 0.00064$} \\[2pt]
$100\theta{}_{\rm s}$ & $1.04190\pm 0.00030$ & $1.04203\pm 0.00030$ \\
 & {\color{blue}$1.04189\pm 0.00029$} & {\color{blue}$1.04211\pm 0.00027$} \\[2pt]
$\ln10^{10}A_{\rm s}$ & $3.041\pm 0.016$ & $3.048\pm 0.015$ \\
 & {\color{blue}$3.043\pm 0.014$} & {\color{blue}$3.053\pm 0.015$} \\[2pt]
$n_{\rm s}$ & $0.9670\pm 0.0041$ & $0.9688\pm 0.0038$ \\
 & {\color{blue}$0.9658\pm 0.0041$} & {\color{blue}$0.9712\pm 0.0032$} \\[2pt]
$\tau{}_{\rm reio}$ & $0.0533\pm 0.0077$ & $0.0577\pm0.0076$ \\
 & {\color{blue}$0.0539\pm 0.0075$} & {\color{blue}$0.0602\pm 0.0074$} \\[2pt]
$\Omega_1$ & $-0.41\pm 0.13$ & $-0.112^{+0.063}_{-0.048}$ \\
 & {\color{blue}$0$} & {\color{blue}$0$} \\[2pt]
$\Omega_2$ & $< 0.0460$ & $0.010^{+0.002}_{-0.010}$ \\
 & {\color{blue}$0$} & {\color{blue}$0$} \\[2pt]
\hline
$H_0$ ($\mathrm{km\ s^{-1}Mpc^{-1}}$) & $75.4^{+3.9}_{-2.3}$ & $69.74\pm 0.77$ \\
 & {\color{blue}$67.48\pm 0.55$} & {\color{blue}$68.44\pm0.28$} \\[2pt]
$w_{\rm de}(0)$ & $-1.36^{+0.24}_{-0.09}$ & $-1.053^{+0.034}_{-0.024}  $ \\
 & {\color{blue}$-1$} & {\color{blue}$-1$} \\[2pt]
$z_{\rm reio}$ & $7.50\pm 0.79$ & $7.96\pm 0.76$ \\
 & {\color{blue}$7.62\pm 0.74$} & {\color{blue}$8.19\pm 0.72$} \\[2pt]
$\Omega_{\rm m}$ & $0.252^{+0.014}_{-0.028}$ & $0.2917\pm 0.0060$ \\
 & {\color{blue}$0.3139\pm 0.0074$} & {\color{blue}$0.3009\pm 0.0036$} \\[2pt]
$\sigma_8$ & $0.879^{+0.032}_{-0.024}$ & $0.825\pm 0.012$ \\
 & {\color{blue}$0.8101\pm 0.0056$} & {\color{blue}$0.8080\pm 0.0060$} \\[2pt]
$S_8$ & $0.803^{+0.013}_{-0.017}$ & $0.8136\pm 0.0094$ \\
 & {\color{blue}$0.829\pm 0.013$} & {\color{blue}$0.8092\pm 0.0083$} \\[2pt]
$r_{\rm d}$ (Mpc) & $147.19\pm 0.28$ & $147.32\pm 0.22$ \\
 & {\color{blue}$147.09\pm 0.27$} & {\color{blue}$147.50\pm 0.20$} \\[2pt]
$z_{\rm d}$ & $1060.04\pm 0.30$ & $1060.09\pm 0.28$ \\
 & {\color{blue}$1060.04\pm 0.28$} & {\color{blue}$1060.20\pm 0.29$} \\[2pt]
$\rm Age$ (Gyr) & $13.617^{+0.079}_{-0.110}$ & $13.736\pm 0.024$ \\
 & {\color{blue}$13.790\pm 0.023$} & {\color{blue}$13.758\pm 0.017$} \\[2pt]
\hline
$\rm \Delta \chi^2_{\text{min}}$ & $-3.08$ & $-4.08$ \\
\hline
\end{tabular}
\end{table}

\begin{figure}[h]
    \centering
    \includegraphics[width=8.5cm]{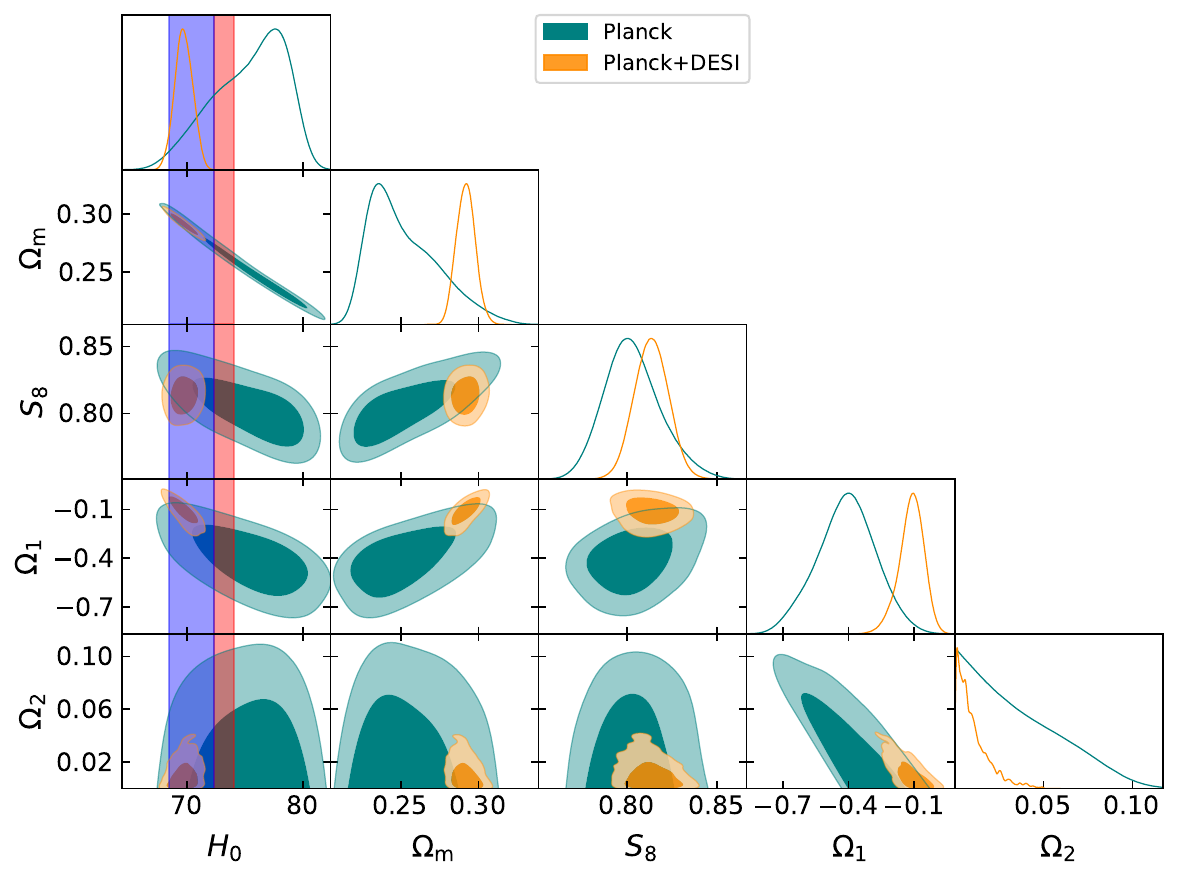}
        \caption{Marginalized 1D posteriors and joint 2D contours (68\% and 95\% CL) for some derived parameters of $\Omega_1\Omega_2-\Lambda$CDM model using Planck (Teal) and Planck+DESI (Orange) datasets. Vertical band on left (Blue) stands for $H_0=70.39\pm1.93$ km s$^{-1}$Mpc$^{-1}$ (CCHP TRGB measurement \cite{Freedman:2019jwv}), and on right (Red) for $H_0=73.17\pm0.86$ km s$^{-1}$Mpc$^{-1}$ (SH0ES measurement \cite{Breuval:2024lsv}).}
    \label{fig1}
\end{figure} 

\begin{figure}[h]
    \centering
    \includegraphics[width=7.1cm]{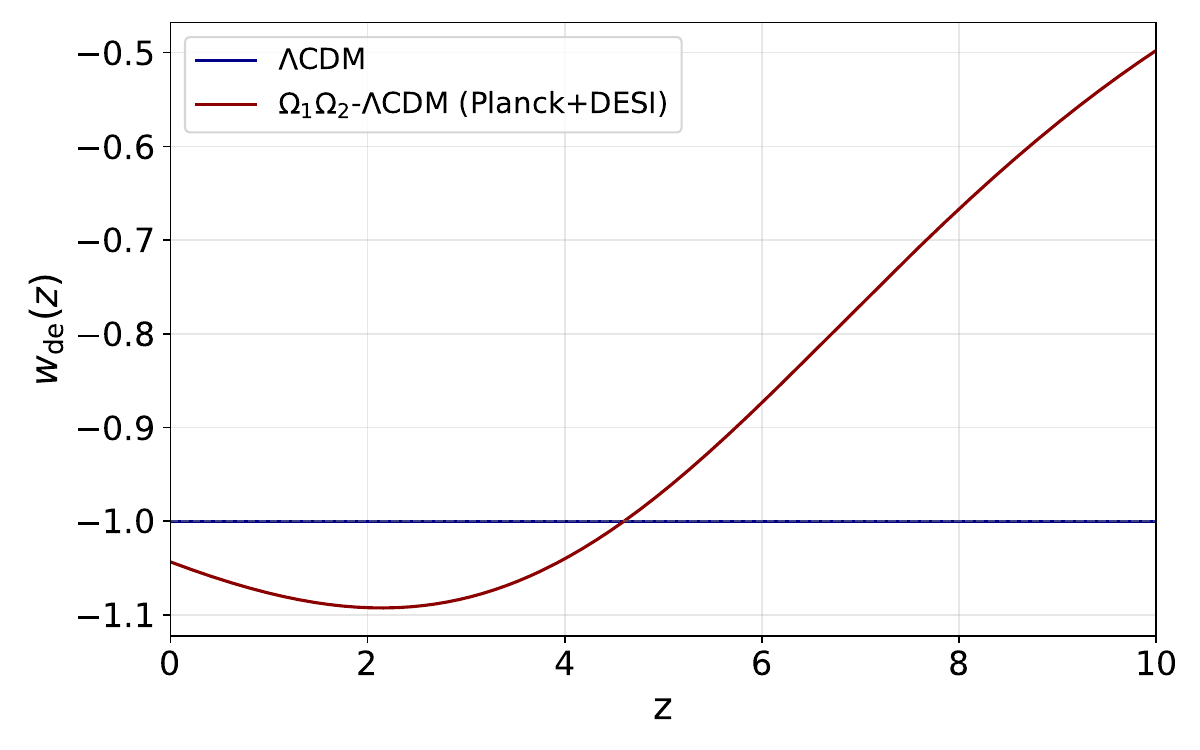}
    \includegraphics[width=7.1cm]{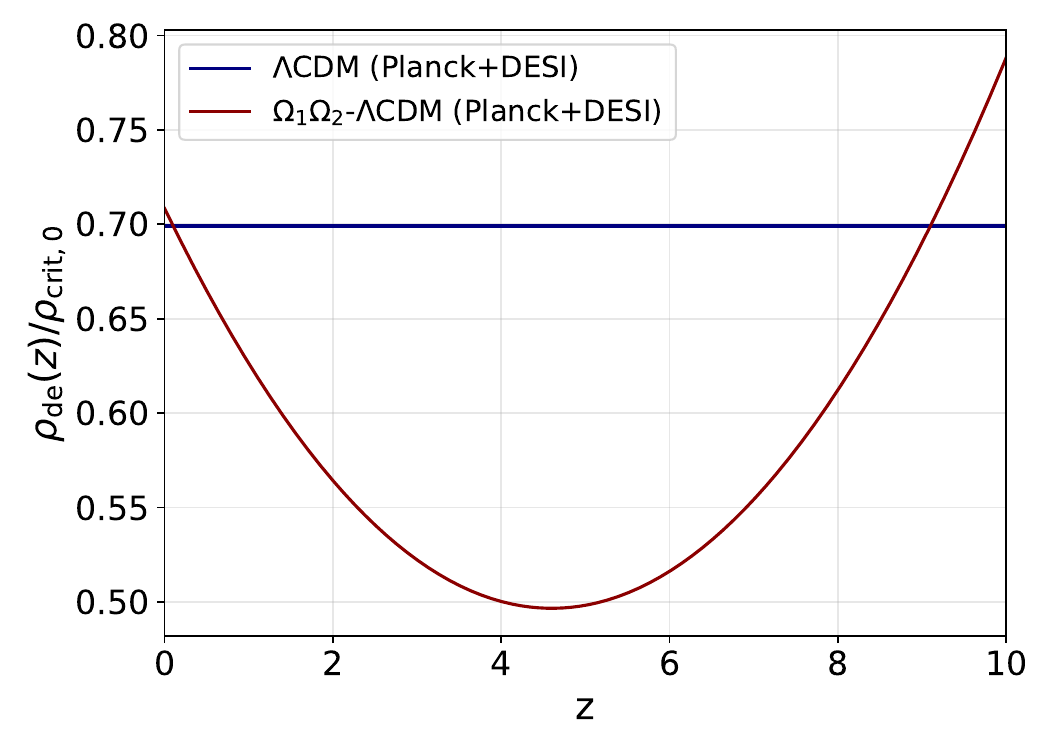}
    \includegraphics[width=7.1cm]{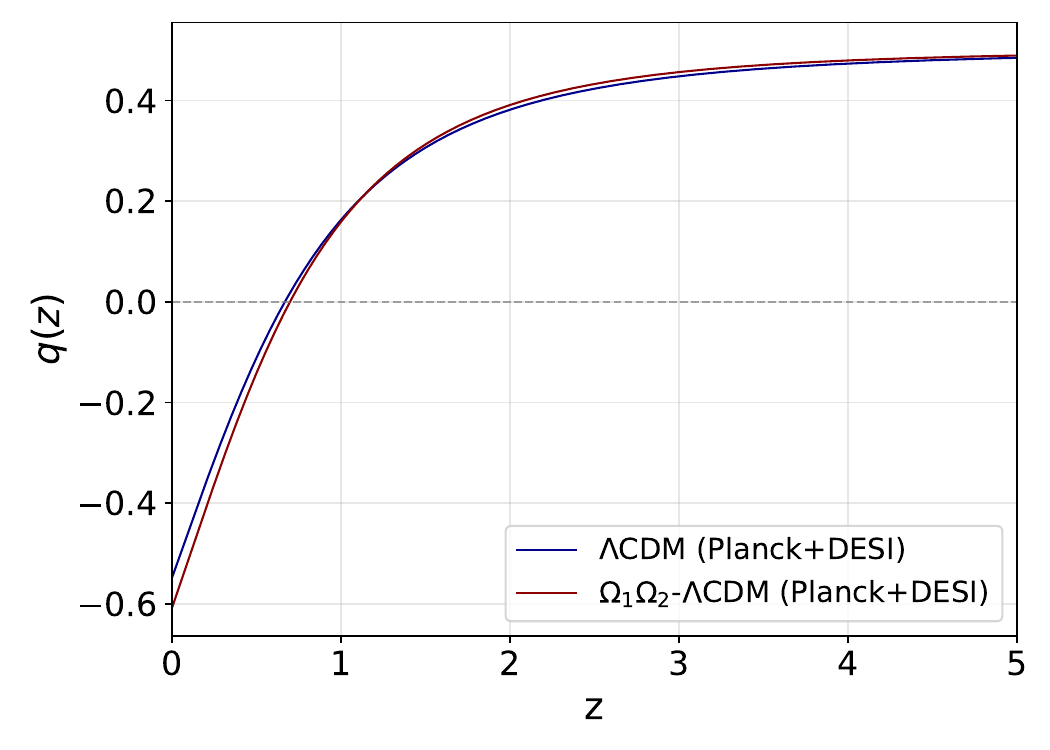}
    \includegraphics[width=7.1cm]{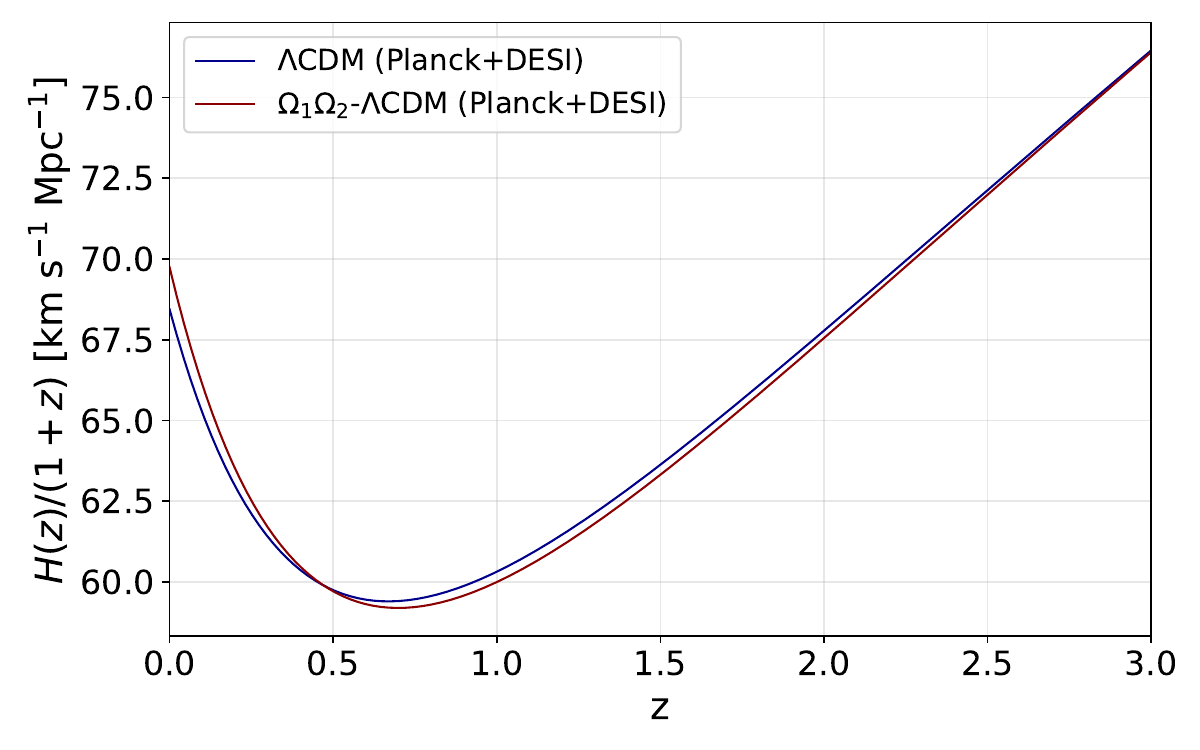}
        \caption{Evolution of $w_{\rm de}(z)$, $\rho_{\rm de}/\rho_{\rm crit 0}$,  $q(z)$ and $H(z)/(1+z)$  versus $z$ for $\Lambda$CDM and $\Omega_1\Omega_2$-$\Lambda$CDM models with mean values of parameters inferred from the Planck+DESI analyses as shown in Table \ref{tab:results}.} 
    \label{fig2}
\end{figure} 

\section{Results and Discussion}\label{sec:results}
In this section, we interpret the parameter constraints summarized in Table~\ref{tab:results}, the marginalized posteriors in Fig.~\ref{fig1} and reconstructed background quantities shown in Fig.~\ref{fig2}. We describe the physical implications of the dynamical dark energy contributions to the total energy density through $(\Omega_1,\Omega_2)$.

When constrained using Planck data alone, the $\Omega_1\Omega_2$--$\Lambda$CDM model shows the characteristic geometric degeneracy for the extended dark-energy scenario under consideration, as expected. As anticipated, the standard early-universe parameters $(\omega_b,\omega_{\rm cdm},n_s,A_s,\tau)$ remain tightly constrained and essentially unchanged relative to $\Lambda$CDM, while the late-time sector shows substantial freedom via the additional free parameters $\Omega_1$ and $\Omega_2$ of dark energy. We notice that the Planck only data allow a significantly negative value of $\Omega_1 = -0.41\pm0.13$, while $\Omega_2$ remains weakly bounded from above. This freedom provides a wide range of the Hubble constant accommodating the CCHP TRGB and SH0ES measurements of $H_0$ (Vertical bands in Fig. \ref{fig1}). This explicitly demonstrates  that the CMB does not intrinsically prefer a low value of $H_0$, once additional late-time degrees of freedom are introduced, for instance in the dark energy sector. Instead, the apparent $H_0$ tension arises from the restrictive late-time dynamics assumed in $\Lambda$CDM framework. The improvement in the goodness of fit, $\Delta\chi^2_{\rm min}=-3.08$ relative to $\Lambda$CDM, indicates that the additional freedom provided by the dynamical dark energy parameters $(\Omega_1,\Omega_2)$ is mildly favored by the CMB, even in the absence of low-redshift data. However, this preference is accompanied by large degeneracies  among the related parameters, leading to broadened posteriors for $H_0$, $\Omega_{\mathrm{m}}$ and $S_8$ (see Fig. \ref{fig1}).

When we include the DESI BAO measurements, which nicely anchor the late-time expansion history in the redshift range $0.295\leq z\leq2.330$, in the analyses, the parameter spaces alter dramatically with improved goodness of fit $\Delta\chi^2_{\rm min}=-4.08$. We observe that the DESI data efficiently break the geometric degeneracy present in the CMB by directly constraining $H(z)$. With Planck+DESI, the allowed range of $\Omega_1$ is significantly reduced to
$\Omega_1 = -0.112^{+0.063}_{-0.048}$,
while $\Omega_2$ exhibits weak but nontrivial constraint around a small positive value, $\Omega_2\sim10^{-2}$, possibly due to subdominance of the term $\Omega_2(1+z)^2$ in the effective redshift range $0.295\leq z\leq2.330$ of the DESI BAO measurements. Overall, we notice that the DESI BAO dataset captures dynamical dark energy deviations from a pure cosmological constant provided by the $\Omega_1\Omega_2$--$\Lambda$CDM framework. Correspondingly, the inferred Hubble constant shifts to $H_0 = 69.74\pm0.77~\mathrm{km\;s^{-1}Mpc^{-1}}$
which is in excellent agreement with the CCHP TRGB measurements: $H_0=69.80\pm1.9$ km s$^{-1}$Mpc$^{-1}$  \cite{Freedman:2019jwv} and the latest best (highest precision) estimate $H_0 = 70.39 \pm 1.22$ km s$^{-1}$Mpc$^{-1}$ \cite{Freedman:2024eph}.  Importantly, this relaxation of $H_0$ tension is achieved without altering early-universe physics, as evidenced by the near-identical constraints on $r_{\rm d}$ and $z_{\rm d}$ in $\Omega_1\Omega_2$--$\Lambda$CDM relative to the $\Lambda$CDM model. Furthermore, we find $\Omega_1=0.010^{+0.031}_{-0.011}$ (99\% CL), which confirms dynamical dark energy at $\sim 3\sigma$ level in line with the recent DESI DR2 observations \cite{DESI:2025zgx}.


The reconstructed evolution of the dark energy equation of state $w_{\rm de}(z)$ along with $\rho_{\rm de}(z)/\rho_{\rm crit 0}$,  $q(z)$ and $H(z)/(1+z)$, as shown in Fig.~\ref{fig2}, clearly demonstrates a non-monotonic behavior of dark energy that clearly distinguishes the $\Omega_1\Omega_2$--$\Lambda$CDM model from the standard $\Lambda$CDM. This evolution shows a physically meaningful transition between quintessence-like ($w_{\rm de}>-1$) and phantom-like ($w_{\rm de}<-1$) regimes, with the characteristic transitions at specific redshifts. At high redshifts ($z\gtrsim5$), $w_{\rm de}(z)$ stays above $-1$, indicating a quintessence-like behavior where dark energy dilutes with the expnasion of universe but remains subdominant to matter and radiation. This regime preserves standard early-universe physics and explains why CMB observables remain largely insensitive to the $\Omega_1$ and $\Omega_2$ parameters. Later, the equation of state smoothly crosses the phantom divide $w_{\rm de}=-1$ at $z = 4.6$, and enters a phantom phase. This transition results apparently from the competing effects of the negative $\Omega_1(1+z)$ and positive $\Omega_2(1+z)^2$ terms, where the later becomes dynamically important enough to slow the dilution of dark energy. The smooth and continuous crossing indicates that the model realizes effective phantom behavior without having any singularities or instabilities at the background level. Following this crossing, $w_{\rm de}(z)$ decreases further, reaching a minimum value of $w_{\rm de} = -1.09$ at $z = 2.14$. This marks a distinct phantom epoch where dark energy density grows relative to a cosmological constant. Physically, this enhances late-time acceleration and leads to a faster expansion rate compared to $\Lambda$CDM, naturally yielding higher inferred values of $H_0$ in $\Omega_1\Omega_2$--$\Lambda$CDM framework, while preserving early-time observables and the sound horizon scale. At lower redshifts ($z \lesssim 1$), the equation of state exhibits reverse trend, increasing back toward $w_{\rm de}=-1$. This indicates that the phantom behavior is transient and not persistent. The dark energy equation of state continues to grow towards $-1$ with $w_{\rm de} = -1.05$ at the present epoch, avoiding extreme future scenarios like a Big Rip. The overall evolution characterized by quintessence-like ($w_{\rm de}>-1$) at high-$z$, phantom crossing at $z=4.6$, phantom phase with minimum $w_{\rm min}=-1.09$ at $z=2.14$, and return toward $w=-1$ at present, reflects a nicely balanced dark energy sector where $\Omega_1$ and $\Omega_2$ contributions dominate at different epochs. Thus, The reconstructed equation of state exhibits a transition from a quintessence-like phase
(\(w_{\rm de} > -1\)) to a phantom phase (\(w_{\rm de} < -1\)), followed by an asymptotic
approach to \(w_{\rm de} = -1\) at late times. This behavior is characteristic of
well-behaved Quintom dark energy models with a de~Sitter attractor
\cite{Feng:2004ad,Guo:2004fq,Cai:2009zp}. The corresponding effects of this dynamical behavior of dark energy are also visible in the evolution of $\rho_{\rm de}(z)/\rho_{\rm crit 0}$,  $q(z)$ and $H(z)/(1+z)$ (see Fig.~\ref{fig2}). Notice that the effective energy density of dark energy does not become negative despite being negative contribution from the second term $\Omega_1(1+z)^2$. The plot of $q(z)$ shows that the onset of acceleration takes place early in $\Omega_1\Omega_2$--$\Lambda$CDM compared to $\Lambda$CDM.  In summary, the early quintessence behavior preserves standard early-time cosmology, the intermediate phantom phase boosts late-time acceleration and relaxes the Hubble tension, and the final approach toward $w=-1$ stabilizes the present universe, gradually diminishing the effect of transient phantom-phase. 

The transient phantom phase in the dark energy equation of state enhances late-time acceleration (raising $H_0$) while suppressing structure growth, leading to the matter clustering parameter $S_8 = 0.8136\pm0.0094$ (from Planck+DESI) which are in closer agreement with reported constraints from weak lensing surveys, such as recent analyses of KiDS-Legacy, $S_8 = 0.815^{+0.016}_{-0.021}$ \cite{Wright:2025xka}, compared to standard $\Lambda$CDM. While this improved concordance is promising, we point out that these weak lensing datasets were not included in our analysis; direct fitting to these datasets is required to properly assess whether the $\Omega_1\Omega_2$--$\Lambda$CDM model does not have the $S_8$ tension.

Collectively, the constraints reported here provide a consistent physical scenario where the $\Omega_1\Omega_2-\Lambda$ dark energy sector introduces a transient modification to the late-time expansion history of universe. This modification leaves the early-universe observables in tact and improves concordance with low-redshift measurements of the expansion rate. The correlated shifts across multiple parameters suggest the observational data considered here respond to genuine dynamical dark energy features rather than statistical noise. Future investigations incorporating growth-sensitive probes like redshift-space distortions and cosmic shear tomography will be essential to further constrain the $\Omega_1\Omega_2-\Lambda$CDM model, particularly the $\Omega_2$ parameter.

\section{Comparison with some Dynamical Dark Energy Models}
\label{sec:comparison}

While our framework is mathematically principled and physically interpretable, it is important to situate it within the existing landscape of other existing dynamical dark energy models providing extensions of $\Lambda$CDM framework, especially the widely-used Chevallier--Polarski--Linder (CPL) parameterization \cite{Linder:2006xb} and other recent Taylor-expansion approaches \cite{Cheng:2025lod}. We show that the $\Omega_1\Omega_2$-$\Lambda$CDM model yields qualitatively different and observationally preferred behavior, particularly in light of the observational data from Planck and DESI.

At low redshift $(|z| \ll 1)$, we can expand Eq.~\eqref{eq:weff} around $z=0$ to obtain
\begin{equation}
w_{\rm de}(z) \approx w_0 + w_a \left( \frac{z}{1+z} \right) + \mathcal{O}(z^2),
\label{eq:cpl_limit}
\end{equation}
where 
\begin{align}
w_0 &= -1 + \frac{1}{3} \frac{\Omega_1 + 2\Omega_2}{\Omega_{\Lambda} + \Omega_1 + \Omega_2}, \\
w_a &= \frac{1}{3} \frac{\Omega_{\Lambda}(\Omega_1 + 4\Omega_2) - \Omega_1\Omega_2}{(\Omega_{\Lambda} + \Omega_1 + \Omega_2)^2}.
\end{align}
This recovers the approximate form of the CPL parametrization, demonstrating that our model encompasses this widely-used ansatz as a low-redshift limit. However, unlike CPL, our model provides a \textit{well-defined} evolution at all redshifts $z \in [-1, \infty)$, with explicit conditions ensuring non-singular behavior. Further, it provides higher order approximation compared to CPL while keeping the same number of free parameters in the Friedmann equation.

From the analyses of $\Lambda$CDM and CPL models with different datasets, the DESI collaboration \cite{DESI:2025zgx} reported that $\Lambda$CDM is being
challenged by the combination of DESI BAO with other measurements unless there is an
unknown systematic error associated with one or more datasets and that dynamical dark
energy provides a possible solution, which emphasizes the need to study dynamical dark energy models. The constraints on the CPL model with Planck+DESI data reported by DESI collaboration are:
$H_0 = 63.6^{+1.6}_{-2.1} \; \mathrm{km\,s^{-1}\,Mpc^{-1}}$,
    $\Omega_m = 0.353 \pm 0.021$,
    $w_0 = -0.42 \pm 0.21$,
    and $w_a = -1.75 \pm 0.58$. These results obviously worsen the Hubble tension compared to \(\Lambda\)CDM and imply a strongly phantom equation of state $(w \ll -1)$ at high redshifts, which is theoretically unstable and inconsistent with early-universe constraints.

In contrast, the \(\Omega_1\Omega_2\)-\(\Lambda\)CDM model with the same number of parameters yields:
$H_0 = 69.74 \pm 0.77 \; \mathrm{km\,s^{-1}\,Mpc^{-1}}$, $\Omega_m = 0.2917 \pm 0.0060$, and a well-behaved $w_{\rm de}(z)$ that respects $w_{\rm de}(z) \geq -1$ at high $z$, as shown in Figure \ref{fig2}. The model is in excellent agreement with the independent TRGB determination $H_0 = 70.39 \pm 1.22 \; \mathrm{km\,s^{-1}\,Mpc^{-1}}$ and preserves the success of early-universe cosmology. Moreover, within this framework, DESI BAO data prefer a dynamical dark energy component at $\sim 3\sigma$ relaxing the Hubble tension, while CPL also predicts dynamical behavior of dark energy but it often exhibits overfitting without improving concordance, worsening the $H_0$ tension by breaking the degeneracies in an opposite direction compared to $\Omega_1\Omega_2$-$\Lambda$CDM model, as may be observed in the results on CPL reported by DESI Collaboration  \cite{DESI:2025zgx}.

In a recent study  \cite{Cheng:2025lod}, the authors propose a phenomenological dark energy model based on a Taylor expansion of the pressure $p(a)$ in powers of $(1-a)$, leading to a dynamical equation of state $w(a)$. While they report a strong statistical preference for dynamical dark energy $(>4\sigma$ in their second-order model with Planck+DESI+DESY5), we note a critical contrast with our own model.
Their \textit{second-order} model, when constrained with Planck+DESI alone, yields $H_0 = 62.8^{+1.4}_{-2.2} \; \mathrm{km\,s^{-1}\,Mpc^{-1}}$, which is \textit{lower} than the Planck $\Lambda$CDM baseline and \textit{worsens} the Hubble tension with both SH0ES and TRGB measurements. Even in their best-fit combination (CMB+DESI+DESY5, $\Delta\chi^2 = -18.41)$, they find $H_0 = 66.73 \pm 0.57$, which remains in significant tension with local determinations ($\sim 3\sigma$ from TRGB). This demonstrates that a mere statistical improvement over \(\Lambda\)CDM does \textit{not} necessarily translate into a mitigation of the key cosmological tensions or an improvement of cosmological concordance.

In stark contrast, the $\Omega_1\Omega_2$-$\Lambda$CDM model---also a second-order expansion, but in $(1+z)$ around the de Sitter future---produces with Planck+DESI alone a Hubble constant of $H_0 = 69.74 \pm 0.77 \; \mathrm{km\,s^{-1}\,Mpc^{-1}}$, in excellent agreement with the CCHP TRGB value. This is achieved with the same observational datasets and a similar number of parameters, yet through a fundamentally different and physically better-motivated expansion scheme. This comparison shows that not all Taylor-expanded phenomenological extensions of dark energy are equivalent. While both works explore Taylor-expanded dark energy, the choice of expansion variable, $(1-a)$ of pressure versus $(1+z)$ of energy density, and the physical interpretation of the terms lead to dramatically different late-time behaviors and cosmological implications. 

We acknowledge that the improvement in $\chi^2$ relative to \(\Lambda\)CDM is modest ($\Delta\chi^2 = -4.08$ for Planck+DESI). However, the primary aim of this work is not to achieve a better statistical fit, but to demonstrate that a minimal, well-motivated extension like $\Omega_1\Omega_2$-$\Lambda$CDM can restore multi-epoch consistency across:
(i) Early universe: Planck CMB $(z \sim 1100)$ (ii) Intermediate redshifts: DESI BAO $(0.3 \lesssim z \lesssim 2.3)$
(iii) Late universe: CCHP TRGB or SH0ES $(z \approx 0)$. The fact that our model yields $H_0$ in agreement with TRGB while preserving early-universe parameters unchanged from $\Lambda$CDM is a non-trivial result. It suggests that the Hubble tension may be relaxed through a controlled, transient modification of the late-time expansion history, without invoking new early-universe physics. Thus, our model not only fits the data but also restores concordance across early, intermediate, and late-universe probes---an outcome that the CPL and pressure-based parameterizations fail to achieve. This highlights the unique value and novelty of our approach, which should be of considerable interest to the cosmology community.

\section{Conclusions}
\label{sec:conclusions}

We have investigated a phenomenological extension of the prevailing standard cosmological framework, the $\Omega_1\Omega_2$--$\Lambda$CDM model, where the total energy density of the universe is parameterized in powers of $1+z$, leading to a physically interpretable dynamics of effective dark energy. The model nicely captures the late-time deviations from a pure cosmological constant regime.    Our results show that the CMB alone does not intrinsically favor a low Hubble constant; instead, the apparent $H_0$ tension emerges from the restrictive late-time dynamics assumed in the $\Lambda$CDM framework. Planck data within the $\Omega_1\Omega_2$--$\Lambda$CDM framework provide substantial freedom in late-time expansion as evidenced by sufficiently broad degenracies of the dark energy parameters $\Omega_1$, $\Omega_2$, allowing wide range of $H_0$ accommodating TRGB and SH0ES measurements. The inclusion of DESI BAO data breaks the geometric degeneracies, constraining the deviations more tightly and providing $H_0 = 69.74\pm0.77$ km s$^{-1}$ Mpc$^{-1}$, which is in excellent agreement with the recent CCHP TRGB measurements. It should not be considered as a resolution of $H_0$ tension in general. However, we observe that the $\Omega_1\Omega_2$--$\Lambda$CDM model  exhibits multi-epoch consistency at high redshifts (CMB), intermediate redshifts (DESI) and present-day epoch (CCHP TRGB). This concordance indicates that the inferred expansion history in the $\Omega_1\Omega_2$--$\Lambda$CDM model is internally consistent across a wide redshift range, while any remaining discrepancy with Cepheid-based SH0ES measurements may possibly be confined to local distance-ladder systematics rather than global cosmology.

The reconstructed evolution of the equation of state of dark energy in $\Omega_1\Omega_2$--$\Lambda$CDM model for Planck+DESI data reveals a physically rich behavior of dynamical dark energy: starting from quintessence-like behavior at high redshifts, smoothly crossing the phantom divide at $z \simeq 4.6$, reaching a minimum $w_{\rm de} = -1.09$ at $z = 2.14$, and approaching $w_{\rm de} = -1.05$ at present. Overall, it indicates a transient phantom-like regime of dark energy in the late universe, which injects additional dark energy in the late universe, thereby enhancing the expansion rate of the universe, and thereby relaxing the $H_0$ tension. This non-monotonic evolution of the equation of state shows a genuinely dynamical dark energy sector that remains subdominant during early epochs but becomes consequential at late times in a controlled manner. While the reconstructed dark energy equation of state exhibits a smooth crossing of the phantom divide and a transient phantom phase at intermediate redshifts, and this behavior is continuous at the background level, we stress that phantom crossing is generically nontrivial at the perturbative level. The present analysis focuses on background expansion and linear growth constraints and does not address the full dynamical consistency of the model. Establishing perturbative stability, including ghost freedom, positive sound speed, and the absence of strong coupling near the crossing, requires an explicit dynamical realization, such as multi-field or Quintom-like constructions  \cite{Vikman:2004dc,Copeland:2006wr,Feng:2004ad,Guo:2004fq,Cai:2009zp}, which will be investigated in a future work.

In the present work, we have focused mainly on exploring the fundamental characteristics of the $\Omega_1\Omega_2$--$\Lambda$CDM model beginning with the minimal Planck CMB and DESI BAO data. The current evidence with these datasets does not justify a strong statistical preference over $\Lambda$CDM; however, the model highlights interesting directions which merit further exploration. Future work will focus on a detailed perturbation and stability analysis, broader dataset combinations including Type~Ia supernovae, and systematic comparisons with the alternative dark energy models existing in the literature. Such studies will be essential to determine whether the $\Omega_1\Omega_2$--$\Lambda$CDM framework is physically viable and observationally distinguishable. Moreover, the observational data from ongoing and future surveys with larger redshift coverage and statistical precision  may decisively test whether the mild deviations identified here represent genuine new physics or residual statistical fluctuations. On the other hand, a transient phantom phase at low redshift could leave imprints on the late-time expansion history and the structure growth of the universe, which may possibly be probed through complementary astronomical observations, especially the redshift range $z\sim 1-2$ is of particular interest, as it encompasses the peak of cosmic star-formation activity and a major transition in galaxy evolution (``cosmic noon''~\cite{Hopkins:2006bw,Tomczak:2013bxa,Madau:2014bja,Akarsu:2025nns}). Nonetheless, the $\Omega_1\Omega_2$--$\Lambda$CDM framework offers a flexible pathway toward resolving cosmological tensions and advancing concordance cosmology.

\acknowledgments
The author sincerely thanks the referee for valuable comments and constructive suggestions. The author is also grateful to Abraão J. S. Capistrano and Rafael C. Nunes for fruitful discussions. This work was supported by the Startup Research Grant from Plaksha University (File No. OOR/PU-SRG/2023-24/08).

\bibliography{main}

\end{document}